\documentclass[letterpaper,twoside,twocolumn,english,superscriptaddress,showpacs]{revtex4}
\usepackage[T1]{fontenc}
\usepackage[latin9]{inputenc}
\usepackage{babel}
\usepackage{bm}
\usepackage{amsmath}
\usepackage{amssymb}
\usepackage{graphicx}
\usepackage{esint}
\usepackage{color}
\usepackage[unicode=true,pdfusetitle,
 bookmarks=false,
 breaklinks=false,pdfborder={0 0 1},backref=false,colorlinks=false]
 {hyperref}

\makeatletter

\@ifundefined{textcolor}{}
{
 \definecolor{BLACK}{gray}{0}
 \definecolor{WHITE}{gray}{1}
 \definecolor{RED}{rgb}{1,0,0}
 \definecolor{GREEN}{rgb}{0,1,0}
 \definecolor{BLUE}{rgb}{0,0,1}
 \definecolor{CYAN}{cmyk}{1,0,0,0}
 \definecolor{MAGENTA}{cmyk}{0,1,0,0}
 \definecolor{YELLOW}{cmyk}{0,0,1,0}
 }

\makeatother

\begin{document}

\title{Density functional theory of Composite Fermions}

\author{Zhang Yin-Han}
\thanks{email:zhangyinhan2008@gmail.com}

\affiliation{Beijing National Laboratory for Condensed Matter Physics and Institute
of Physics, Chinese Academy of Sciences, Beijing 100190, China}

\author{Shi Jun-Ren}
\thanks{email:junrenshi@pku.edu.cn(Corresponding Author)}
\affiliation{International Center for Quantum Materials, Peking University, Beijing
100871, China}
\affiliation{Collaborative Innovation Center of Quantum Matter, Beijing,100871, China}

\begin{abstract}
We construct a density functional theory for two-dimension electron (hole) gases subjected to both strong magnetic fields and external potentials.  In particular, we are focused on regimes near even-denominator filling factors, in which the systems form composite fermion liquids.  Our theory provides a systematic and rigorous approach to determine the properties of ground states in fractional quantum Hall regime modified by artificial structures. We also propose a practical way to construct approximated functional. 
\end{abstract}


\maketitle
High mobility semiconductor two-dimensional electron (hole) gas (2DEG/2DHG) is one of the cleanest condensed matter systems and hosts some of most important discoveries in physics, including quantum Hall effect~\cite{Klitzing1980} and fractional quantum Hall effect~\cite{Tsui1982}.  Recently, there is a surge of interest in designing and imposing artificial structures in these systems.  For instance, it is proposed that a periodic potential modulation could modify 2DEGs to create graphene-like electronic structure,  upon which one could engineer exotic phases such as topological insulator and superconductivity~\cite{Polini2013}. This has become practical as semiconductor nano-fabrication techniques provide versatile and feasible tools to realize these designs in laboratory~\cite{Evers2013}. Until recently, most of studies along the line are focused on non-interacting systems.  On the other hand, a recent theoretical study reveals that the artificial structure could induce novel and interesting topological phases in strongly-correlated regime of fractional quantum Hall effect (FQHE), namely, the quantum anomalous Hall insulator of composite fermions~\cite{Zhang2014}. The development asks for a systematic approach to treat the extremely difficult theoretical problem involving external magnetic fields, potentials, as well as electron-electron interaction.

Density functional theory provides such a systematic and rigorous approach to tangle the problem.  For the usual solid state systems, the density functional theory has achieved great successes in predicting structures and properties. In the presence of the external magnetic field, the theory needs to be modified and extended to the spin and current-density functional theory (SCDFT)~\cite{Vignale1987,Vignalebook}.  In principle, the theory is rigorous and sufficient to predict ground states in arbitrarily strong magnetic field, provided one knows \emph{exact} energy functional in terms of electron/spin density and paramagnetic current density. In practice, however, the energy functional is determined by the assumption that the paramagnetic current density is small, effectively limiting the applicability of the theory in the regime of small magnetic fields. The more ambitious attempt to predict fractional quantum Hall states within the density functional theory requires substantial modification of the theoretical framework and yields limited success~\cite{Heinonen1995}.  It is clear that to treat our problem, a new approach is needed to circumvent these difficulties.

In this Letter, we develop a density functional theory for composite fermions (CF).  In the theory, instead of using the electron paramagnetic current density, we express the energy functional in term of the paramagnetic current density of \emph{composite fermions}~\cite{Jains,Heinonen,Kalmeyer1992,Willett1993}. It is known that 2DEG/2DHG systems with even-denominator filling factors form composite fermion liquids subjected to zero effective magnetic field, and the CF paramagnetic current density vanishes~\cite{HLR}.  As a result, we are able to construct the energy functional in the vicinity of these filling factors by assuming the smallness of the CF paramagnetic current density.  To do this, we re-formulate the SCDFT for CFs, and present details of derivation for the theoretical framework sketched in an earlier publication~\cite{Zhang2014}. We further determine the gauge invariance of the exchange correlation energy functional and show a practical way to determine it.
  
Consider a spinless 2DEG system near a magnetic filling factor $\nu=1/2p$ with $p$ being an integer under a set of the external scalar field $U(\bm r)$ and vector field $\bm A(\bm r)$. The Hamiltonian for electrons can be written as 
\begin{align}
\hat{H}_{e}=\hat{T}_{e}+\hat V_{ext}+\hat{V}_{ee}
\end{align}
where the kinetic energy and the external scalar potential operators are  
\begin{align}
\hat{T}_{e}=&\sum_{j} \left[ \frac{1}{2m_{b}}\left(-i\hbar \nabla_{j}-\frac{e}{c}\bm A(\bm r_{j})\right)^{2}\right], \\
\hat{V}_{ext}&=\sum_{j}U(\bm r_{j}),
\end{align}
and $\hat{V}_{ee}$ is the electron-electron interaction. For simplicity, we further decompose the vector potential $\bm A$ into $\bm A_{0} + \bm A^\prime$, where $\bm A_0$ is the vector potential that gives rise to a filling factor $\nu=1/2p$, i.e., $B_0 \equiv \mathrm{curl}\bm A_0 = 2p\phi_0 n_0$, with $\phi_0$ being the magnetic flux quantum and $n_0$ being average electron density of the system.

We can reformulate the problem by making a Chern-Simons transformation:
\begin{align}
\Psi(\bm r_{1},..,\bm r_{N}) \longrightarrow \exp(-2pi\sum_{i<j}\arg(\bm r_{i}-\bm r_{j}))\Psi(\bm r_{1},...,\bm r_{N}),
\end{align}  
where $\arg(\bm r)$ is the angle between the vector $\bm r$ and the positive real axis. As a result, the kinetic energy operator is transformed to,  
\begin{align}
\hat{T}_{CS}=&\sum_{j}\left [\frac{1}{2m_{b}} \left(\hat{\bm p}_{j}-\frac{e}{c}(\bm A_{0}(\bm r_{j})+\bm A^\prime(\bm r_{j})+\bm a_{CS}(\bm r_{j})) \right)^{2}\right], \label{Tcs}
\end{align}
and the Chern-Simons vector potential is,
\begin{align}
\bm a_{CS}(\bm r_j)=\frac{p\phi_0}{\pi}\sum_{i\ne j} \frac{\hat{z}\times(\bm r_{j}- \bm r_i)}{|\bm r_{j}- \bm r_i|^2}.
\end{align}

The resulting Hamiltonian can be solved by a mean field approximation $\mathrm {curl}\bm a_{CS}(\bm r) = -2p\phi_0 n(\bm r)$.  With the approximation and in the uniform density limit,  $\bm a_{CS}$ in Eq.~(\ref{Tcs}) will exactly cancel $\bm A_0$.  It gives rise to the very successful CF picture of the fractional quantum Hall effect, in which each electron attaches $2p$ quantum vortices to form a CF subjected to an effective vector potential $\bm A^\prime(\bm r)$ of reduced effective magnetic strength~\cite{Jains,HLR,Heinonen,Kalmeyer1992,Willett1993}. In particular, systems with $\nu=1/2p$ is equivalent to a CF liquid with zero effective magnetic field~\cite{HLR}.
   
Starting with the CF representation, we can develop a density functional theory to go beyond the usual CF mean field theory~\cite{Zhang2014,HLR,Kalmeyer1992,Heinonen}. We introduce the CF-paramagnetic current density:   
\begin{align}
\bm j_p(\bm r_1) &= \int \mathrm{d}\bm r_2 \dots \mathrm{d}\bm r_N \left[
-\frac{i\hbar}{2m_b}\left(\Psi^* (\bm\nabla_1 \Psi) - (\bm\nabla_1 \Psi^*)  \Psi\right) \right.\notag\\ 
&\left. -\frac{e}{c}\left(A_0(\bm r_1) +\frac{p\phi_0}{m_b\pi}\sum_{i\ne 1} \frac{\hat{z}\times(\bm r_1 - \bm r_i)}{|\bm r_1 - \bm r_i|^2}\right) |\Psi|^2 \right]
\end{align}
It is important to note that $\bm j_p(\bm r)$ vanishes for a uniform 2DEG/2DHG at $\nu=1/2p$.

We can prove generalized Hohenberg-Kohn (HK) theorem which dictates that ground state energy is a functional of density $n(\bm r)$ and CF-paramagnetic current density $\bm j_p(\bm r)$. First, we prove that the ground-state distributions of $n(\bm r)$ and  $\bm j_{p}(\bm r)$ uniquely determine the external scale potential $U(\bm r)$ and vector potential $\bm A^\prime (\bm r)$, and hence the non-degenerate ground state $|\Psi \rangle$. To see this, we assume that there exist two sets of external fields $U_{1(2)}(\bm r)$, $\bm A^\prime_{1(2)}(\bm r)$ that yield the same set of distributions $n(\bm {r})$ and $\bm j_{p}(\bm r)$. The CF Hamiltonian $\hat{H}_{1(2)}$ have ground-state $|\Psi_{1(2)}\rangle$ with their ground-state energy $E_{1(2)}$. For $|\Psi_{1}\rangle$ and $\hat{H}_{1}$, we have
\begin{align} 
&E_{1}<\langle \Psi_{2}|\hat{H}_{1}|\Psi_{2}\rangle=E_{2}+\int d^{2} r (U_{2}(\bm r)-U_{1}(\bm r))n(\bm r) \notag \\ 
&-\frac{e}{c} \int d^{2} r (\bm A_{1}(\bm r)-\bm A_{2}(\bm r))\cdot \bm{j}_{p}(\bm r)\notag\\
&+\frac{e^{2}}{2m_{b}c^2} \int d^{2} r n(\bm r)\left[ \bm A_{1}^{2}(\bm r)-\bm A_{2}^{2}(\bm r)\right]. \notag      
\end{align}
Similarly, for $|\Psi_{2}\rangle$ and $\hat{H}_{2}$, another inequality is obtained by exchanging the suffix $(1\leftrightarrow2)$. Summing up the two inequalities, one observes contradiction  $E_{1}+E_{2}<E_{2}+E_{1}$. 

Since the non-degenerate ground state wave function $|\Psi\rangle$ is a functional of the density distributions $n(\bm r)$ and $\bm j_{p}(\bm r)$, the energy expectation value can also be expressed as a functional: 
\begin{align}
E_{U,\bm A^{\prime}}[n,\bm j_{p}]=F[n,\bm j_{p}]+\int d^2 r n(\bm r) U(\bm r)\notag \\ 
-\frac{e}{c}\int d^2 r \bm A^{\prime}(\bm r)\cdot \bm j_{p}(\bm r)+\frac{e^{2}}{2m_{b}c^{2}} \int d^2 r \bm A^{\prime 2}(\bm r)n(\bm r)
\label{EVJ}
\end{align}
with a universal functional defined as
\begin{equation}
F[n,\bm j_{p}]\equiv\min_{\Psi^{\prime}\rightarrow \{n,\bm j_{p}\}}\langle \Psi^{\prime}|\hat{T}_0+\hat{V}_{ee}|\Psi^{\prime}\rangle
\label{UniversalF}
\end{equation}
where $\hat{T}_0 \equiv \left. \hat{T}_{CS}\right|_{\bm A^\prime = 0}$, and the search for the minimum is restricted in all possible asymmetry wave function $\Psi^{\prime}$ that yield the give density distributions.  With the energy functional, the ground state density distributions can be obtained by minimizing the functional against $n(\bm r)$ and $\bm j_p(\bm r)$~\cite{Vignalebook}.

We can further map the many-body problem to an effective non-interacting one.  To do this, we assume that for any ground state density distribution yielded by the interacting system, one can find a non-interacting reference system that has the same set of ground-state density distributions. The non-interacting system is obtained by solving Kohn-Sham equation:
\begin{align}
\left [\frac{1}{2m_b}(-i\hbar \nabla -\frac{e}{c}\bm A_{KS}(\bm r))^{2}+U_{KS}(\bm r)\right] \psi_{j}(\bm r)=\epsilon_{j}\psi_{j}(\bm r) 
\label{KSEQ}
\end{align}
And the corresponding density distributions are determined by,
\begin{align} 
\bm j_{p}(\bm r)&=\sum^{N}_{j=1}\frac{\hbar}{2m_{b}i}\left(\psi^{\ast}_{j}(\bm r)\nabla \psi_{j}(\bm r)-c.c \right)\notag\\
&-\frac{e}{m_{b}c} \left(\bm {A}_{0}(\bm r)+\bar{\bm {a}}_{CS}(\bm r)\right)n(\bm r), \notag \\
& n(\bm r)=\sum_{i=1}^{N}|\psi_{i}|^2. 
\end{align} 
where the summation is for the $N$-lowest energy states, and
\begin{equation}
\bar{\bm a}_{CS} (\bm r)= \frac{p\phi_0}{\pi}\int \mathrm{d} \bm r^\prime \frac{\hat{z}\times(\bm r- \bm r^\prime)}{|\bm r- \bm r^\prime|^2} n(\bm r^\prime),
\end{equation}
is the mean field expectation value of $\bm a_{CS}$.  In Eq.~(\ref{KSEQ}), we introduce Kohn-Sham potentials $U_{KS}$ and $\bm A_{KS}$, which are functionals of the density distributions.

For the non-interacting system, there is also a HK theorem dictating the existence of a universal energy functional:
\begin{align}
T_s =\min_{\Psi_{s}\rightarrow \{n,\bm j_{p}\}}\langle \Psi_{s}|\hat{T}_{0}|\Psi_{s}\rangle,
\end{align}
where the search for the minimum is restricted in all possible non-interacting wave functions of slater determinants.
We can decompose the universal energy functional of the interacting system $F[n,\bm j_{p}]$ into:
\begin{align}
F[n,\bm j_{p}]=T_{s}[n,\bm j_{p}]+\frac{1}{2}\int d^2 r_{1} r_{2} & n(\bm r_{1})V_{ee}(\bm r)n(\bm r_{2})\notag \\
&+E_{xc}[n,\bm j_{p}],
\label{UF}
\end{align}
where we introduce exchange-correlation energy functional $E_{xc}$ to register residual energy correction.

Using definition of $T_s$, one can re-express it as:
\begin{align}
T_s[n,\bm j_{p}]=\sum_{i}\epsilon_{i}-\frac{e}{c}\int d^{2} r \left[ \bm A_{0}(\bm r)+\bm a_{CS}(\bm r)-\bm A_{KS}(\bm r) \right]\notag \\
\cdot \left[\bm j_{p}(\bm r)+\frac{e}{m_{b}c}n(\bm r)(\bm A_{0}(\bm r)+\bm a_{CS}(\bm r))\right]-n(\bm r)U_{KS}(\bm r)\notag\\
+\int d^{2} r n(\bm r)\left[\frac{e^{2}}{2m_{b}c^{2}}\left((\bm A_{0}(\bm r)+\bm a_{CS}(\bm r))^{2}-\bm A^{2}_{KS}(\bm r) \right)\right],
\label{Ts}
\end{align}
with $\epsilon_{j}$, $\bm a_{CS}(\bm r)$, and KS potentials $U_{KS}$ and $\bm A_{KS}$ being functionals of $n$ and $\bm j_{p}$.

Substituting Eq.~(\ref{Ts}) into Eq.~(\ref{UF}) and then Eq.~(\ref{EVJ}), we obtain the total energy functional $E_{U,\bm A^{\prime}}[n,\bm j_{p}]$. To determine KS potentials, we carry out minimization of energy functional $E_{U,\bm A^{\prime}}[n,\bm j_{p}]$ with respect of $n(\bm r)$ and $\bm j_{p}(\bm r)$:
\begin{align}
\frac{\delta E_{U,\bm A^{\prime}}[n,\bm j_{p}]}{\delta n(\bm r)}=0, \quad \frac{\delta E_{U,\bm A^{\prime}}[n,\bm j_{p}]}{\delta \bm j_{p}(\bm r)}=0. 
\end{align}

The KS potentials can then be determined:  
\begin{align}
&\bm A_{KS}(\bm r)=\bm A(\bm r)+\bm a_{CS}(\bm r)+\bm A_{xc}(\bm r),\notag\\
&U_{KS}(\bm r)=U(\bm r)+U_{xc}(\bm r)+\int dr^{\prime} V_{ee}(\bm r, \bm r^{\prime})n(\bm r^{\prime})  \notag\\
&+\frac{e^{2}}{2m_{b}c^{2}}\left(\bm A^{\prime2}(\bm r)-(\bm A^\prime(\bm r)+\bm A_{xc}(\bm r))^{2}\right)+2p\phi_{0}m_{z}(\bm r),
\end{align}
where $m_{z}(\bm r)$ is the orbital magnetization density of the non-interacting KS system~\cite{Shi2007}, and the exchange-correlation corrections to the KS potentials are: 
\begin{align}
U_{xc}(\bm r)\equiv\frac{\delta E_{xc}[n,\bm j_{p}]}{\delta n(\bm r)}, \quad \bm A_{xc}(\bm r)\equiv-\frac{c}{e}\frac{\delta E_{xc}[n,\bm j_{p}]}{\delta \bm j_{p}(\bm r)}.
\label{EXD} 
\end{align}

We can show that gauge invariance puts important constraint on the possible form of exchange-correlation energy functional. We introduce the gauge transformation:
\begin{align}
\bm A(\bm r) \rightarrow \bm A_{G}=\bm A(\bm r)+\nabla f(\bm r),
\end{align}
with an arbitrary analytic function $f(\bm r)$.  The density distributions $n$ and $\bm j_{ p}$, and the wave function $\Psi$ of the system are transformed as 
\begin{align}
&n_{G}(\bm r) =n(\bm r), \quad \Psi_{G}=\exp(\sum_{j}\frac{ief(\bm r_{j})}{\hbar c})\Psi,  \notag\\
&\bm j_{Gp}(\bm r) =\bm j_{p}(\bm r)+\frac{e}{m_{b}c}\nabla f(\bm r)n(\bm r).
\end{align}
The invariance of the energy functional under the gauge transformation dictates:
\begin{align}
F&[n,\bm j_{p}+\frac{e}{m_{b}c}n\nabla f]=F[n,\bm j_{p}]\notag\\
&+\frac{e^{2}}{2m_{b}c^{2}}\int d^2 r n(\bm r)(\nabla f(\bm r))^{2}+\frac{e}{c}\int d^{2}r \nabla f(\bm r)\cdot \bm j_{p}(\bm r).
\end{align}
The kinetic functional $T_{s}$ for the non-interacting KS system transforms similarly.  It follows that the exchange-correlation functional is gauge invariant: 
\begin{align}
E_{xc}[n(\bm r),\bm j_{p}(\bm r)+\frac{e}{m_{b}c}n(\bm r) \nabla f(\bm r)]=E_{xc}[n(\bm r),\bm j_{p}(\bm r)].
\label{EX}
\end{align}

As a result, the exchange-correlation energy functional can be rewritten as,
\begin{align}
E_{xc}[n,\bm j_{p}]\equiv \bar{E}_{xc}[n,\bm v],
\end{align}
which only depends on vorticity defined as:
\begin{align} 
\bm v(\bm r)\equiv \nabla \times \frac{\bm j_{p}(\bm r)}{n(\bm r)}.
\end{align}
The exchange-correlation scalar and vector potential can be rewritten as 
\begin{align}
U_{xc}(\bm{r}) =\frac{\delta\bar{E}_{xc}[n,\bm{\nu}]}{\delta n(\bm{r})}+&\frac{e}{c}\bm{A}_{xc}(\bm r)\cdot\frac{\bm{j}_{p}(\bm{r})}{n(\bm{r})}, \notag \\
-\frac{e}{c}\bm{A}_{xc}(\bm{r}) =-\frac{1}{n(\bm{r})}&\nabla\times\frac{\delta\bar{E}_{xc}[n,\bm{\nu}]}{\delta\bm{\nu}(\bm{r})}.
\end{align} 

To determine $E_{xc}$, we employ the local density approximation (LDA) by assuming that $E_{xc}$ at position $\bm r$ is only determined by the local density $n(\bm r)$.  We further exploit the fact that the CF paramagnetic current density vanishes at the uniform limit and at $\nu=1/2p$.  It suggests that $\bm j_p(\bm r)$ should be small in the vicinity of even-denominator filling factors.  Based on the stiffness theorem~\cite{Vignalebook}, the universal functional $F$ can be approximated as, in the uniform density limit: 
\begin{align}
F[n,\bm j_{p}]=F[n, 0]-\frac{1}{2 V}\sum_{\bm q}\delta\bm j_{p}(-\bm q) \chi^{-1}_{\bm j_{p},\bm j_{p}}(\bm q)\delta\bm j_{p}(\bm q)
\end{align}    
where the static paramagnetic current response function matrix $\chi_{\bm j_{p},\bm j_{p}}(\bm q)$ is defined as 
\begin{align}
\chi_{\bm j_{p},\bm j_{p}}(\bm q)=-i\int dt \Theta(t)\langle [\bm j_{p}(\bm q,t),\bm j_{p}(-\bm q,0)]\rangle 
\end{align}
For an isotropic system, we can decompose
\begin{align}
\chi_{\bm j_{p},\bm j_{p}}(\bm q)=\chi_{\parallel}(\bm q)\frac{q_{i}q_{j}}{q^{2}}+\chi_{\perp}(\bm q)(\delta_{i,j}-\frac{q_{i}q_{j}}{q^{2}})
\end{align}
into a longitudinal component $\chi_{\parallel}(\bm q)$ and a transverse component $\chi_{\perp}(\bm q)$. Because a purely longitudinal and static vector potential cannot induce any physical current, i.e., $\chi_{\parallel}(\bm q, 0)=-n/m$,  we can consider transverse current response $\chi_{\perp}(\bm q)$ only. 

It is straightforward to determine the exchange-correlation energy functional:
\begin{align}
E_{xc}[n,\bm j_{p}]=&-\frac{1}{2V}\sum_{\bm q}\delta\bm j^{\perp}_{p}(-\bm q)(\chi^{-1}_{\perp}(\bm q)-\chi^{(0)-1}_{\perp}(\bm q))\delta\bm j^{\perp}_{p}(\bm q)\notag \\
&+E_{xc}[n,0],
\label{ec}
\end{align}
where $\chi_{\perp}^{0}$ is the transverse response function for the non-interacting system.

The exchange-correlation energy functional has the general form:
\begin{align}
&\bar{E}_{xc}[n,\bm v]=E_{xc}[n,0]+\frac{1}{2}\int d r^{2}_1 d r^{2}_{2} f(\bm r_{1} -\bm r_{2}) \bm v(\bm r_{1}) \cdot \bm v(\bm r_{2})
\label{EEX}
\end{align}
where the function $f(\bm r)$ is related to the transverse current response function by, 
\begin{align}
f(\bm r)=\notag 
 \frac{1}{V}\sum_{\bm q} \frac{n^{2}}{q^{2}}(\chi^{-1}_{\perp}(\bm q)-\chi^{(0)-1}_{\perp}(\bm q))\exp(i\bm q\cdot \bm r).
\end{align}
In general, $f(\bm r)$ is a function of the density $n$, which can be replaced with average density $n_0$, as the first order of approximation. 

We have formulated the current density functional theory for CFs. A gauge invariant self-consistent Kohn-Sham equation for CFs is constructed. The general form of the exchange-correlation functional is established. Explicit determination of the exchange-correlation functional requires numerical evaluation of ground state energy of uniform 2DEG at  $\nu = 1/2p$ and corresponding transverse current response function, which will be investigated in the future publications.   

This work is supported by 973 program of China (2012CB921304) and National Science Foundation of China (11325416).

\end{document}